\documentclass[aps,pra,showpacs,twocolumn]{revtex4}
\usepackage{amssymb}
\usepackage{amsmath}
\usepackage{graphicx}
\usepackage{epsfig}

\begin{document}

\title{Quantum State Transfer Characterized by Mode Entanglement}
\author{Xiao-Feng Qian$^{1,2}$, Ying Li$^{1}$, Yong Li$^{2}$, Z. Song$^{1,a}$
and C. P. Sun$^{1,2,a,b}$}
\affiliation{$^{1}$Department of Physics, Nankai University, Tianjin 300071, China}
\affiliation{$^{2}$ Institute of Theoretical Physics, Chinese Academy of Sciences,
Beijing, 100080, China}

\begin{abstract}
We study the quantum state transfer (QST) of a class of
tight-bonding Bloch electron systems with mirror symmetry by
considering the mode entanglement. Some rigorous results are
obtained to reveal the intrinsic relationship between the fidelity
of QST and the mirror mode concurrence (MMC), which is defined to
measure the mode entanglement with a certain spatial symmetry and
is just the overlap of a proper wave function with its mirror
image. A complementarity is discovered as the maximum fidelity is
accompanied by a minimum of MMC. And at the instant, which is just
half of the characteristic time required to accomplish a perfect
QST, the MMC can reach its maximum value one. A large class of
perfect QST models with a certain spectrum structure are
discovered to support our analytical results.
\end{abstract}

\maketitle

\section{Introduction}

Quantum entanglement is a fascinating feature of quantum theory of many body
systems \cite{chung}. The concurrence \cite{Wootters:98}, as a widely used
measure of pairwise entanglement defined for the spin-1/2 systems, has been
intensively investigated. Through various concurrences defined by different
authors, people have explored the relations between entanglement and some
physical observables such as energy and momentum etc. \cite{entangled rings,
Qian}, as well as the relations between entanglement and some physical
phenomena, such as quantum correlation \cite{Wang} and quantum phase
transitions etc. \cite{Osterloh:02, Lin, Chen}.

On the other hand, people have proposed many protocols for the quantum state
transfer (QST) recently \cite{Bose, Christandal, Shi,Li,kaba}. In these
schemes based on quantum spin systems, almost without any spatial or
dynamical control over the interactions among qubits, the quantum state can
be transferred with high fidelity through a quantum channel, or quantum data
bus, which is necessary for scalable quantum computations based on realistic
silicon devices. The physical process of QST through a quantum spin system
can be understood as a dynamical permutation or translation preserving the
initial shape of a quantum state, which can be realized as a specific
evolution of the total quantum spin system from an initial wave function
localized around a single site of the lattice to a distant one. The basic
feature of QST is characterized by fidelity, which is usually the overlap of
the identical image of an initial state with its transferred counterpart.

This paper will be devoted to understand the intrinsic relation between
quantum entanglement and QST for the engineered quantum spin chains, or
quantitatively, between concurrence and fidelity. Some rigorous results are
obtained to reveal the essential relationship between these two fascinating
issues for the tight-bonding Bloch electrons. Actually, the QST from one
location to another can be considered as perfect if the fidelity can reach
its maximum value one at some instants. Literally, the perfect QST is a
dynamic process starting from an initially factorized state (product state)
to a finally factorized state through a middle process with the
superposition of factorized states. Since a superposition of single particle
states of Bloch electrons can be understood as a mode entanglement \cite{x
wang}, the studies of QST can be naturally referred to the various phenomena
of quantum entanglement.

Motivated by arguments about the entanglement concurrence and the quantum
correlations \cite{Wootters:98,x wang}, \ we first define the mirror mode
concurrence (MMC) $C(t)$ to characterize the mode entanglement of a wave
packet in Bloch electron systems with mirror symmetry. It will be proved
that the MMC is no less than the overlap of the wave packet at time $t$ with
its mirror image. By quantitatively comparing the MMC with the time
dependent fidelity $F(t)$ of QST, a novel complementary relation is
discovered as the increase of $F(t)$ is accompanied by a decrease of $C(t)$
(vice versa). Especially, at the instant $\tau /2,$ where $\tau $ is the
characteristic time to accomplish a perfect QST with $F(\tau )=1$, the MMC
can reach its maximum
\begin{equation}
C(\tau /2)=\max (C(t))=1.
\end{equation}%
An engineered Bloch electron model with a certain spectrum structure, which
admits perfect QST, is discovered and used to demonstrate this complementary
relation through numerical simulations.

\section{One-dimensional Bloch electron system with mirror symmetry}

We consider a one-dimensional Bloch electron system in an engineered crystal
lattice of $N$ sites with mirror symmetry with respect to the center of the
lattice. The model Hamiltonian with tight-bonding approximation is written
as
\begin{equation}
H=\sum_{j=1}^{N-1}J_{j}a_{j}^{\dag }a_{j+1}+h.c.  \label{H}
\end{equation}%
in terms of the fermion creation (annihilation) operator $a_{j}^{\dag }$ ($%
a_{j}$), where the site-dependent coupling constants $J_{j}$ are real. The
single-particle space is spanned by $N$ basis vectors%
\begin{eqnarray}
&&\left\vert 1\right\rangle =\left\vert 1\text{, }0\text{, }0\text{, }...%
\text{, }0\text{, }0\right\rangle ,  \notag \\
&&\left\vert 2\right\rangle =\left\vert 0\text{, }1\text{, }0\text{, }...%
\text{, }0\text{, }0\right\rangle ,  \notag \\
&&......................., \\
&&\left\vert N\right\rangle =\left\vert 0\text{, }0\text{, }0\text{, }...%
\text{, }0\text{, }1\right\rangle  \notag
\end{eqnarray}%
where $\left\vert n_{1}\text{, }n_{2}\text{, }...\text{, }n_{N}\right\rangle
$ ($n_{j}=0$, $1$) denotes the Fock state of fermion systems. Then the
reflection operator $R$ is defined as $R|j\rangle =|N+1-j\rangle $.
Obviously, the mirror symmetry by $[R,H]=0$ means that $J_{j}=J_{N-j}$ .

We describe the localized electron state around the $l$-th site as a
superposition $\left\vert \psi _{l}\right\rangle =\sum_{j}c_{j}\left\vert
j\right\rangle $ with the summation over the small domain containing the
site $l$. This assumption means that $\left\vert \psi _{l}\right\rangle $ is
a wave packet around the site $l$. If $l^{\prime }$ denotes another site far
from the site $l$, we can approximately assume the vanishing overlap $%
\left\langle \psi _{l}\right. \left\vert \psi _{l^{\prime }}\right\rangle
\simeq 0$ for two wave packets $\left\vert \psi _{l}\right\rangle $ and $%
\left\vert \psi _{l^{\prime }}\right\rangle $. With this assumption the
perfect QST is described as the dynamic process that the initial state $%
\left\vert \psi _{l}\right\rangle $ can evolve exactly into its mirror
image. Mathematically, the time evolution operator $U(t)=\exp (-iHt)$
becomes the reflection operator $R$ at the instant $\tau $, i.e., $U(\tau
)=R $. We define the fidelity as
\begin{equation}
F_{j}(t)=\left\vert \left\langle R\psi _{j}\right\vert U(t)\left\vert \psi
_{j}\right\rangle \right\vert =\left\vert \left\langle \psi _{j}\right\vert
R^{\dagger }U(t)\left\vert \psi _{j}\right\rangle \right\vert .
\end{equation}%
A perfect QST can be depicted by the maximized fidelity $F_{j}(\tau )=1$.

Now we can intuitionally recognize that QST phenomenon is associated with
the mode entanglement.\ In the terminology of mode entanglement, the single
electron state
\begin{eqnarray}
\left\vert E\right\rangle &=&\alpha \left\vert 1\right\rangle +\beta
\left\vert N\right\rangle \\
&\equiv &\alpha \left\vert 1,0,...,0\right\rangle +\beta \left\vert
0,0,...,1\right\rangle  \notag
\end{eqnarray}%
can be regarded as an entangled state if the single fermion at the $1$-th
site and $N$-th site can be probed in principle \cite{x wang}. In this sense
$\left\vert \psi _{l}\right\rangle $ can be viewed as an $N$-component
entanglement. The perfect QST from $\left\vert 1\right\rangle $ to $%
\left\vert N\right\rangle $ through the middle state $\left\vert \psi
(t)\right\rangle =U(t)\left\vert 1\right\rangle $ can be understood as a
dynamic process starts from a localized (unentangled) state $\left\vert
1\right\rangle $ to another localized state $\left\vert N\right\rangle $
through the entangled state $\left\vert \psi (t)\right\rangle $.

\section{Mirror mode concurrence as the fingerprints of perfect QST}

Actually a QST is a process, during which mode entanglement is generated
first and then destroyed. To quantitatively characterize this dynamic
feature, we define the mirror mode concurrence (MMC)
\begin{equation}
C(t)=\sum_{j=1}^{N/2}C_{j,N+1-j}
\end{equation}%
with respect to a pure state, evolved from a localized initial wave packet, $%
\left\vert \psi (t)\right\rangle =U(t)\left\vert \psi (0)\right\rangle $.
Here each term $C_{j,N+1-j}$ in the summation concerns two separated sites,
the site $j$ and its mirror imagine $l=N+1-j,$ \ and is defined by the
pairwise mode concurrence \cite{x wang}
\begin{equation}
C_{jl}=2\max \left\{ 0,\text{ }\left\vert Z_{jl}\right\vert -\sqrt{%
X_{jl}^{+}X_{jl}^{-}}\right\} ,
\end{equation}%
constructed in terms of the correlation functions
\begin{equation}
Z_{jl}=\left\langle a_{j}^{\dagger }a_{l}\right\rangle
,X_{jl}^{+}=\left\langle \hat{n}_{j}\hat{n}_{l}\right\rangle ,
\end{equation}%
and%
\begin{equation}
X_{jl}^{-}=\left\langle (1-\hat{n}_{j})(1-\hat{n}_{l})\right\rangle .
\end{equation}%
where the average $\left\langle {}\right\rangle $ is defined with respect to
the pure state $\left\vert \psi (t)\right\rangle .$

The physical significance\ will be in two folds that explicitly reveals the
close relationship between the mode entanglement and the dispersion of the
wave packet in time evolution. Firstly we notice that $Z_{jl}$ and $%
X_{jl}^{-}$\ are the non-zero elements of the two mode reduced density matrix%
\cite{x wang}
\begin{eqnarray}
\rho _{jl} &=&tr_{N-2}(\left\vert \psi (t)\right\rangle \langle \psi (t)|)
\notag \\
&=&\left(
\begin{array}{cccc}
X_{jl}^{+} &  &  &  \\
& Y_{jl}^{+} & Z_{jl}^{\ast } &  \\
& Z_{jl} & Y_{jl}^{-} &  \\
&  &  & X_{jl}^{-}%
\end{array}%
\right)
\end{eqnarray}%
for a system conserving total particle number, where $tr_{N-2}$ means
tracing over the variables besides the two on the sites $j$ and $l=N+1-j$.
In the single-particle subspace we have $X_{jl}^{+}=0$ and thus
\begin{equation}
C(t)=\sum_{j=1}^{N}\left\vert Z_{j,N+1-j}\right\vert .
\end{equation}%
It is obvious that MMC is a generalization of the usual entanglement
measure--concurrence and thus characterize the quantum entanglement in some
sense.

Secondly the MMC defined above have a geometric interpretation for the
dynamic dispersion of the wave packet. We rewrite the MMC as
\begin{equation}
C(t)=\sum_{j=1}^{N}\left\vert \psi (j,t)\right\vert \left\vert \psi
(N+1-j,t)\right\vert
\end{equation}%
where $\psi (j,t)=\left\langle j\right. \left\vert \psi (t)\right\rangle $.
It is easy to show that \
\begin{eqnarray}
C(t) &\geq &|\sum_{j=1}^{N}\langle \psi (t)\left\vert j\right\rangle
\left\langle j\right\vert R\left\vert \psi (t)\right\rangle |  \notag \\
&=&\left\vert \langle \psi (t)|R\left\vert \psi (t)\right\rangle \right\vert
\end{eqnarray}%
where we have used $RR^{\dagger }=R^{\dagger }R=1$. The above equation
clearly implies that $C(t)$ is no less than the overlap integral of the
state $\left\vert \psi (t)\right\rangle $ with its mirror image. Especially,
for a large class of states $\left\vert \psi (t)\right\rangle
=\sum_{j=1}^{N}c_{j}\left\vert j\right\rangle $ listed \ in two situations
as follows, $C(t)$ is exactly equal to the overlap integral: (i) The
electronic wave function are completely localized in a finite domain $%
D=[1,N/2]$ with no overlap with its mirror image $[N/2,N]$. In this case,
the MMC vanishes exactly. (ii) The the coefficients of each pair of mirror
symmetric non-zero components in $\left\vert \psi (t)\right\rangle $ have
the same sign or opposite sign.
\begin{figure}[tbp]
\includegraphics[angle=-90, bb=311 59 494 595, width=7 cm, clip]{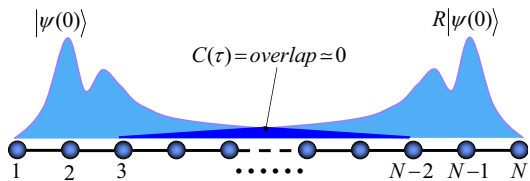}
\caption{(color online) Illustration of the mirror mode concurrence $C(t)$
at $t=\protect\tau $. $C(\protect\tau )$ is just the overlap integral of the
wave function $R\left\vert \protect\psi (0)\right\rangle $ with its mirror
image when $\left\vert \protect\psi (0)\right\rangle $ satisfies either of
the two situations (i) and (ii).}
\end{figure}

For a perfect QST accomplished at the instant $t=\tau $, the evolution
operator $U(\tau )$ becomes the reflection operator $R$ and $\left\vert \psi
(0)\right\rangle $ evolves exactly into its mirror image $R\left\vert \psi
(0)\right\rangle $. Since the initial wave packet $\left\vert \psi
(0)\right\rangle $ is usually a very localized wave function, the wave
function $\left\vert \psi (\tau )\right\rangle =R\left\vert \psi
(0)\right\rangle $ and its mirror image $\left\vert \psi (0)\right\rangle $
almost do not overlap with each other (see the illustration in Fig. 1). Thus
we have
\begin{equation}
C(\tau )=C(0)\geq \left\vert \langle \psi (0)|R\left\vert \psi
(0)\right\rangle \right\vert \simeq 0.
\end{equation}%
Therefore, at $t=\tau $, the MMC $C(\tau )$ almost vanishes when the
fidelity $F(t)$ reaches its maximum ($F(\tau )=1$).

From the above argument we see that there exists a quite interesting
relationship between entanglement and fidelity. We provide a model more
universal than the QST model in Ref. [10]. Their mode is a mapping to the
collective spin system with $SU(2)$ dynamic symmetry (by $J_{j}=J_{0}\sqrt{%
j\left( N-j\right) }$ more concretely), but our model only requires a much
smaller mirror symmetry (by $J_{j}=J_{N-j}$ more generally ) and thus have
much\ wider applications. In fact we have shown many examples in Ref. \cite%
{Li} as well as in the following discussions. The further arguments about
the complementarity relationship between entanglement and fidelity will also
be\ presented in such a general framework.

\section{Maximal mode entanglement}

The above analysis has confirmed our intuition about the complementary
relation between the fidelity of QST and the MMC of mode entanglement. As
for the other feature of this complementary relation, we need to consider
when the MMC can reach its maximum.

Obviously there exists the inequality
\begin{equation}
C(t)\leq \frac{1}{2}\sum_{j=1}^{N}(\left\vert \psi (j,t)\right\vert
^{2}+\left\vert \psi (N+1-j,t)\right\vert ^{2})=1,
\end{equation}%
which takes the equal sign only when the wave function evolves into its
mirror imagine, i.e.,
\begin{equation}
\left\vert \psi (j,t)\right\vert =\left\vert \psi (N+1-j,t)\right\vert
\label{eq1}
\end{equation}%
at some instants $t$. This means that $C(t)$ will reach its maximum $\max
(C(t))=1$ at the instants when Eq. (\ref{eq1}) holds.

In order to determine the time when $C(t)$ reaches its maximum one, we need
to solve the equation (\ref{eq1}) about time $t$. To this end we use a
time-independent real symmetric matrix $W$ to diagonalize the Hamiltonian $H$
or the evolution operator $U(t)$ as $WU^{\dagger }(t)W^{T}=A(t)$, where $%
A(t) $ is a diagonal matrix. With these notations, the above equation (\ref%
{eq1}) can be transformed into
\begin{equation}
\left\vert \left\langle \psi _{W}\right\vert A(t)\left\vert
W_{j}\right\rangle \right\vert =\left\vert \left\langle \psi _{W}\right\vert
Q(t)\left\vert W_{j}\right\rangle \right\vert ,  \label{eq2}
\end{equation}%
where%
\begin{eqnarray}
\left\vert \psi _{W}\right\rangle &=&W\left\vert \psi (0)\right\rangle ,
\notag \\
\left\vert W_{j}\right\rangle &=&W\left\vert j\right\rangle , \\
Q(t) &=&WU^{\dagger }(t)U(\tau )W^{T},  \notag
\end{eqnarray}%
We notice that, in general, $\left\vert W_{j}\right\rangle =W\left\vert
j\right\rangle $. $\left\vert \psi _{W}\right\rangle $ and $\left\vert
W_{j}\right\rangle $ are real for a real initial state $\left\vert \psi
(0)\right\rangle $. Then the solutions to the equation (\ref{eq2}) are
sufficiently\ given by $Q(t)=A^{\dagger }(t)$ or $Q(t)=A(t)$, of which the
non-trivial one is just $t=\tau /2$. Indeed, \ since $\ \left\langle \alpha
\right\vert A\left\vert \beta \right\rangle =\left\langle \beta \right\vert
A\left\vert \alpha \right\rangle $ for any two real vectors $\left\vert
\alpha \right\rangle $ and $\left\vert \beta \right\rangle $, we have
\begin{eqnarray}
\left\vert \left\langle \psi _{W}\right\vert Q(t)\left\vert
W_{j}\right\rangle \right\vert &=&\left\vert \left\langle \psi
_{W}\right\vert A^{\dagger }(t)\left\vert W_{j}\right\rangle \right\vert \\
&=&\left\vert \left\langle W_{j}\right\vert A^{\dagger }(t)\left\vert \psi
_{W}\right\rangle \right\vert =\left\vert \left\langle \psi _{W}\right\vert
A(t)\left\vert W_{j}\right\rangle \right\vert .  \notag
\end{eqnarray}%
Therefore, the solution $t=\tau /2$ is obviously given by $Q(t)=A^{\dagger
}(t)$ or%
\begin{equation}
U^{\dagger }(t)U(\tau )=U(t).
\end{equation}

We summarize the above argument as a proposition: If $F(t)${\ reach its
maximum }$1$ at the instant $t=\tau ${, then at time }$t=\tau /2$,{\ }$%
C(\tau /2)=1$. In appendix A, we will prove its inverse proposition: if $C(t)$%
{\ reach its maximum }$1$ at the instant $t=\tau /2${, then at
time }$t=\tau $,{\ }$F(\tau )=1$. Furthermore, we can generalize
these conclusion for the more general situation even with a higher
dimensional Hamiltonian (also see the appendix A)

The solution $t=\tau /2$ to the equation (\ref{eq2}) indicates that the time
required to form the maximal mode entanglement is just half of the time
needed to implement the perfect QST. Furthermore we can prove that, for a
real vector $\left\vert \psi (0)\right\rangle $, the MMC $C(t)$ is symmetric
with respect to both $t=\tau /2$ and $t=\tau $, namely,
\begin{eqnarray}
C(\frac{\tau }{2}-t) &=&C(\frac{\tau }{2}+t)\text{,}  \label{a} \\
\text{ }C(\tau -t) &=&C(\tau +t).  \notag
\end{eqnarray}%
Actually, for the second equation in Eqs. (\ref{a}) we have
\begin{equation}
C(\tau \pm t)=\sum_{j=1}^{N}\left\vert \left\langle \psi (0)\right\vert
U_{\pm }(t)R^{\dagger }\left\vert j\right\rangle \right\vert \left\vert
\left\langle \psi (0)\right\vert U_{\pm }(t)\left\vert j\right\rangle
\right\vert ,
\end{equation}%
where $U(t)_{+}=U^{\dagger }(t)$ and $U(t)_{-}=U(t)$. Obviously the second
equation in Eqs. (\ref{a}) holds since we have
\begin{equation}
\left\vert \left\langle \psi (0)\right\vert U(t)V\left\vert j\right\rangle
\right\vert =\left\vert \left\langle \psi (0)\right\vert U^{\dagger
}(t)V\left\vert j\right\rangle \right\vert
\end{equation}%
for $V=1,$ $R^{\dagger }$. Also, the first equation in Eqs. (\ref{a}) will
give a similar proof.
\begin{figure}[tbp]
\includegraphics[bb=11 12 233 176, width=7 cm, clip]{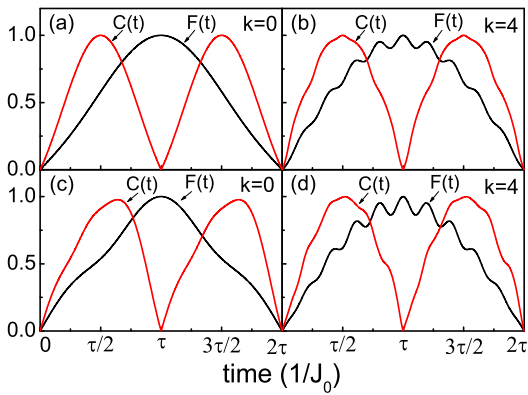}
\caption{(color online) Plots of $C(t)$ and $F(t)$ of the states $\left\vert
\protect\psi (t)\right\rangle $ evolve from real ($(a)$ and $(b)$) and
complex ($(c)$ and $(d)$) initial wave packets $\left\vert \protect\psi %
_{1,2}(0)\right\rangle $. The evolutions of $\left\vert \protect\psi %
(t)\right\rangle $ in $(a)$ and $(c)$ ($(b)$ and $(d)$) are driven by the
4-site Hamiltonian with $k=0$ ($k=4$). For $(a)$ and $(b)$, $F(\protect\tau %
)=1$ and $C(\protect\tau /2)=1$, while for $(c)$ and $(d)$, $F(\protect\tau %
)=1$ and $\max (C(t)),C(\protect\tau /2)\neq 1$.}
\end{figure}

Numerical methods are now employed to give a demonstration of the above
analytical results. We concern a class of schemes that admit perfect QST,
which are presented in Ref. \cite{Shi}. The couplings of the Hamiltonian $H$
are given that%
\begin{equation}
J_{j}=J_{0}\sqrt{(j+\theta _{j}k)\left( N-j+\theta _{j}k\right) }
\end{equation}
where $\theta _{j}=1-(-1)^{j}$, $k=0$, $1$, $2$, $...$ and $J_{0}$ is a
constant. This model possesses a commensurate structure of energy spectrum
that is matched with the corresponding parity. We demonstrate the exact
numerical results of the models with $N=4$ and $k=0$, $4$ in Figs. 2(a) and
2(b). Actually, when $k=0$ (Fig. 2(a)), the model is just the one proposed
in Ref. \cite{Christandal}. We have used the localized initial wave packet
as $\left\vert \psi _{1,2}(0)\right\rangle =c_{1}\left\vert 1\right\rangle
+c_{2}\left\vert 2\right\rangle $, where $c_{1}=5/6$, $c_{2}=\sqrt{11/36}$.
From Figs. 2(a) and 2(b) we can observe that $C(0)=C(\tau )=0$, $C(\tau
/2)=1 $, and $C(t)$ is symmetric with respect to $t=\tau $, $\tau /2$. These
results are in agreement with our analytical results. It also implies the
complementary relation between MMC and fidelity, for inside the range from $%
t=\tau /2$ to $3\tau /2$, the increase of $F(t)$ is accompanied by a
decrease of $C(t)$ (vice versa).

It is pointed out that our results about MMC\ $C(t)$ at $t=\tau /2$ are
based on the condition that the initial wave packet $\left\vert \psi
(0)\right\rangle $ is real except for a global phase. One may be interested
in the situation when $c_{1}$ and $c_{2}$ are not real for $\left\vert \psi
_{1,2}(0)\right\rangle $. For this situation, e.g., $c_{1}=(1+i)/2$, $%
c_{2}=1/5+i\sqrt{23/50}$, the numerical calculation shows that $C(t)$ is not
just symmetric with respect to $t=\tau /2$, $C(\tau /2)\neq 1$, $\max \left(
C(t)\right) $ is very close, yet not equal to one and $C(\tau /2)\neq \max
(C(t))$ (see Figs. 2(c) and 2(d)).

\section{Perfect QST of Bloch electrons in an engineered lattice}

Based on the above recognitions about the relation between a perfect QST and
mode entanglement, we can construct various lattice models with mirror
symmetry to achieve perfect QST. Furthermore we can characterize these QSTs
with the MMC. Actually, a large class of models for QST have been discovered
by us most recently \cite{Shi} by generalizing the spin model in Ref. \cite%
{Christandal}.

Now we further generalize the perfect QST model to a much larger class. The
Hamiltonian is given in Eq. (\ref{H}) with the engineered coupling constants
\begin{equation}
J_{j}=J_{0}\sqrt{(j+\xi _{j})\left( N-j+\xi _{j}\right) },
\end{equation}%
where%
\begin{equation}
\xi _{j}=[1-(-1)^{j}]l/(2m+1)
\end{equation}%
for the given $m,$ $l\in 0,$ $1,$ $2,$ $3$, $...$. We notice that it will
return to the previous models in Refs. \cite{Christandal,Shi} when $m=0$.

Numerical analysis shows that the above Hamiltonian possesses a commensurate
structure of energy spectrum by an experiential formula
\begin{equation}
\varepsilon _{n}=N_{n}E_{0}-(N+1)J_{0},  \label{spectrum}
\end{equation}%
where the energy unit is
\begin{equation}
E_{0}=\frac{2J_{0}}{2m+1},
\end{equation}%
$N_{n}=n(2m+1)-l$ for $n=1$, $2$, $...$, $N/2,$ and $N_{n}=n(2m+1)+l$ for $%
n=N/2+1$, $...$, $N.$ Numerical results show that the above experiential
formula (\ref{spectrum}) still holds when $N=3000$. It can be checked that
the energy spectrum is matched with the corresponding parity (the
eigen-value of $R$) as%
\begin{equation}
p_{n}=(-1)^{N_{n}}\exp \{i[(m+\frac{1}{2})N+1]\pi \}.
\end{equation}%
The corresponding eigen-states $\left\vert \varphi _{n}\right\rangle
=\sum_{j=1}^{N}c_{j}(n)\left\vert j\right\rangle $ can be determined by the
matrix equation $H\left\vert \varphi _{n}\right\rangle =\varepsilon
_{n}\left\vert \varphi _{n}\right\rangle .$

According to Refs. \cite{Shi, Li}, the characteristic time to perform a
perfect QST is $\tau =\pi /E_{0}$, provided that $l/(2m+1)$ is an
irreducible fraction. Now\ we can show that, at $t=\tau $, the time
evolution operator
\begin{equation}
U(t)=\sum_{n}\exp (-i\varepsilon _{n}t)\left\vert \varphi _{n}\right\rangle
\left\langle \varphi _{n}\right\vert
\end{equation}%
is just the mirror reflection operator $R$ by neglecting a global phase,
namely,
\begin{equation}
U(\tau )=\sum_{n}(-1)^{N_{n}}\left\vert \varphi _{n}\right\rangle
\left\langle \varphi _{n}\right\vert =(-1)^{l}R.
\end{equation}%
Thus, the present model admits perfect QSTs when%
\begin{equation}
\xi _{j}=[1-(-1)^{j}]l/(2m+1).
\end{equation}

In order to verify the prediction about the relation between the MMC and
fidelity, a numerical analysis is carried out for the present QST model. We
investigate the 4-site case with $m=1$\ and $l=2$. The real initial wave
packet is also $\left\vert \psi _{1,2}(0)\right\rangle $. Detail behaviors
of the MMC and fidelity between the instants $t=0$ and $t=2\tau $ are shown
in Fig. 3. We notice, in Fig. 3, that $C(t)=0,$ $1,$ $0,$ for $t=0,$ $\tau
/2,$ $\tau $ respectively and $C(t)$ is symmetric with respect to $t=\tau ,$
$\tau /2$. Obviously, it is in agreement with our prediction.
\begin{figure}[tbp]
\includegraphics[ bb=11 11 173 111, width=7 cm, clip]{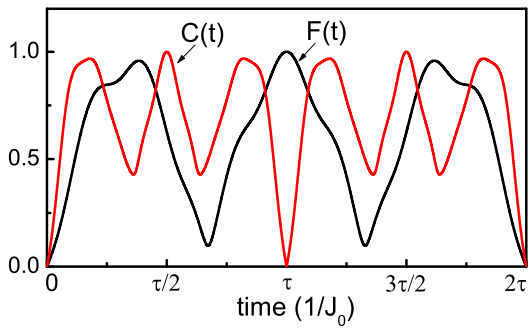}
\caption{(color online) Plots of $C(t)$ and $F(t)$ of the state $\left\vert
\protect\psi (t)\right\rangle =U(t)\left\vert \protect\psi %
_{1,2}(0)\right\rangle $ for the Hamiltonian with $N=4$, $m=1$ and $l=2$. It
shows that $C(0)=C(\protect\tau )=0$, $C(\protect\tau /2)=1$, and $C(t)$ is
symmetric with respect to the $t=\protect\tau ,\protect\tau /2$.}
\end{figure}

\section{Summary}

In summary we have defined the mirror mode concurrence (MMC) to describe how
a perfect quantum state transfer (QST) can be achieved for a large class of
lattice model of fermion systems with mirror symmetry. By investigating the
property of MMC of these perfect QST models, a novel complementary relation
between the MMC and fidelity is revealed. Actually our definition of MMC is
just a part of total concurrence \cite{Qian,yang}. However, when the
symmetry of our systems is taken into consideration, MMC is a better
measurement in characterizing the process of a perfect QST. A new class of
QST models are discovered to support our observations. Therefore, a perfect
QST can now be understood as a process of establishing an entanglement and
then destroying it at the correlated instants. Finally we remark that our
main results are valid in other perfect QST models with general symmetries
such as translation, rotation and etc. It is very interesting to further
investigate the QST vs entanglement relation based on solid state systems
with the symmetries described by point groups or the crystallographic space
groups.

\appendix

\section{A general proof for complementarity $F(\protect\tau %
)=1\Longleftrightarrow C(\protect\tau /2)=1$}

{We have proved that the MMC }$C(t)$ {will reach its maximum }$1$ at the
instant $t=\tau /2$ where $\tau $ is the instant, at which the fidelity $F(t)
$ {reaches its maximum} $F(\tau )=1$. {W}e now prove the inverse
proposition: If $C(t)${\ reaches its maximum }$1$ at the instant $t=\tau /2${%
, then at time }$t=\tau ,$ $F(\tau )=1$. Namely, we have a theorem in a
sufficient and necessary statement

{%
\begin{equation}
F(\tau )=1\Longleftrightarrow C(\tau /2).
\end{equation}%
}

{In this appendix, }we will prove the above theorem for\ a general model
even for higher dimensional fermion systems with a Hamiltonian,

\begin{equation}
H=\sum_{i\neq j}^{N}J_{ij}a_{i}^{\dagger }a_{j}\text{,}  \label{general h}
\end{equation}%
on the one particle Fock space spanned by $N$ basis vectors $\{\left\vert
j\right\rangle \}$, $j=1$, $2$, $3$, $...$, $N$. Suppose the Hamiltonian has
a symmetry $S$ and $[S,H]=0$, and the basis vectors can be decomposed into
two subspaces $\{\left\vert n_{j}\right\rangle $ $|j=1$, $2$, $3$, $...$, $%
N/2\}$ and $\{\left\vert m_{j}\right\rangle $ $|j=1$, $2$, $3$, $...$, $N/2\}
$ such that

\begin{equation}
S\left\vert n_{j}\right\rangle =\left\vert m_{j}\right\rangle \text{, }%
S\left\vert m_{j}\right\rangle =\left\vert n_{j}\right\rangle ,
\end{equation}%
then perfect QST requires that at a certain instant $t=\tau $, $U(\tau
)=\exp (-iH\tau )=S$. The case by Eq. (\ref{general h}) is just a
generalization of the situation of mirror symmetry Hamiltonian.

Through the definition of total concurrence

\begin{equation}
C(t)=\sum_{j}^{N/2}C_{n_{j},m_{j}}=\sum_{j}^{N/2}2\left\vert \left\langle
\psi (t)\right\vert a_{n_{j}}^{\dagger }a_{m_{j}}\left\vert \psi
(t)\right\rangle \right\vert ,
\end{equation}%
we can first prove the proposition from\ $F(\tau )=1$\ to\ $C(\tau /2)=1$.

As for an initial state $\left\vert \psi (0)\right\rangle $, the fidelity of
a state $\left\vert \psi (t)\right\rangle =U(t)\left\vert \psi
(0)\right\rangle $ reads as

\begin{equation}
F(t)=\left\vert \left\langle S\psi (0)\right\vert U(t)\left\vert \psi
(0)\right\rangle \right\vert =\left\vert \left\langle \psi (0)\right\vert
S^{+}U(t)\left\vert \psi (0)\right\rangle \right\vert ,  \label{general f}
\end{equation}%
and a perfect QST at $t=\tau $ can be depicted by the maximized fidelity $%
F(\tau )=1$ when

\begin{equation*}
U(\tau /2)U(\tau /2)=U(\tau )=S
\end{equation*}%
satisfies Eq. (\ref{general h}). We calculate the total concurrence of $%
\left\vert \psi (t)\right\rangle $ as

\begin{eqnarray}
C(t) &=&\sum_{j}^{N/2}2\left\vert \left\langle \psi (0)\right\vert
U^{+}(t)a_{n_{j}}^{\dagger }a_{m_{j}}U(t)\left\vert \psi (0)\right\rangle
\right\vert   \notag \\
&=&\sum_{j=1}^{N/2}2\left\vert \left\langle \psi (0)\right\vert
U^{+}(t)\left\vert n_{j}\right\rangle \left\langle m_{j}\right\vert
SS^{+}U(t)\left\vert \psi (0)\right\rangle \right\vert   \notag \\
&=&\sum_{j=1}^{N/2}2\left\vert \left\langle \psi (0)\right\vert
U^{+}(t)\left\vert n_{j}\right\rangle \right\vert \left\vert \left\langle
n_{j}\right\vert U^{+}(t)\left\vert \psi (0)\right\rangle \right\vert .
\end{eqnarray}%
Then at\ the instant $t=\tau /2$

\begin{equation}
C(\tau /2)=\sum_{j=1}^{N/2}2\left\vert \left\langle \psi (0)\right\vert
U^{+}(\tau /2)\left\vert n_{j}\right\rangle \right\vert \left\vert
\left\langle n_{j}\right\vert U^{+}(\tau /2)\left\vert \psi (0)\right\rangle
\right\vert .
\end{equation}%
For real $\left\vert \psi (0)\right\rangle $ we have $\left\vert
\left\langle \psi (0)\right\vert U^{+}(\tau /2)\left\vert n_{j}\right\rangle
\right\vert =\left\vert \left\langle n_{j}\right\vert U^{+}(\tau
/2)\left\vert \psi (0)\right\rangle \right\vert $ and then
\begin{equation}
C(\tau /2)=\sum_{j=1}^{N/2}2\left\vert \left\langle \psi (0)\right\vert
U^{+}(\tau /2)\left\vert n_{j}\right\rangle \right\vert ^{2}=1.
\end{equation}%
Thus we have

\begin{equation}
\max (C(t))=C(\tau /2)=1.
\end{equation}

Now we prove the proposition from\ $C(\tau /2)=1$\ to{\large \ }$F(\tau )=1$%
. According to Eq. (\ref{general f}), if we require $C(\tau /2)=\mathrm{\max
}(C(t))=1$ at some instant $t=\tau /2$, then

\begin{equation}
\left\vert \left\langle \psi (0)\right\vert U^{+}(\tau /2)\left\vert
n_{j}\right\rangle \right\vert =\left\vert \left\langle m_{j}\right\vert
U(\tau /2)\left\vert \psi (0)\right\rangle \right\vert .
\end{equation}%
Therefore we have%
\begin{equation*}
\left\vert \left\langle \psi (0)\right\vert U^{+}(\tau /2)\left\vert
n_{j}\right\rangle \right\vert =\left\vert \left\langle m_{j}\right\vert
U(\tau /2)\left\vert \psi (0)\right\rangle \right\vert ,
\end{equation*}%
or%
\begin{equation*}
\left\vert \left\langle \psi (0)\right\vert U^{+}(\tau /2)\left\vert
n_{j}\right\rangle \right\vert =\left\vert \left\langle \psi (0)\right\vert
U^{+}(\tau /2)\left\vert m_{j}\right\rangle \right\vert ,
\end{equation*}%
or
\begin{equation*}
\left\vert \left\langle \psi (0)\right\vert U^{+}(\tau /2)\left\vert
n_{j}\right\rangle \right\vert =\left\vert \left\langle \psi (0)\right\vert
U^{+}(\tau /2)S\left\vert n_{j}\right\rangle \right\vert
\end{equation*}%
This means $U^{+}(\tau /2)S=U^{+}(\tau /2)$ or $U^{+}(\tau /2)S=U(\tau /2)$.
It has a trivial solution $S=1$ and an approved  non-trivial solution $%
S=U(\tau )$. With the non-trivial solution $S=U(\tau )$, there will be a
perfect QST, i.e.,%
\begin{eqnarray}
F(\tau ) &=&\left\vert \left\langle S\psi (0)\right\vert U(\tau )\left\vert
\psi (0)\right\rangle \right\vert  \\
&=&\left\vert \left\langle \psi (0)\right\vert S^{+}U(\tau )\left\vert \psi
(0)\right\rangle \right\vert =1.  \notag
\end{eqnarray}%
As is stands, we have verified the theorem $F(\tau
)=1\Longleftrightarrow C(\tau /2)$ in a general situation.

We acknowledge the support of the NSFC (grant No. 90203018, 10474104,
10447133), the Knowledge Innovation Program (KIP) of Chinese Academy of
Sciences, the National Fundamental Research Program of China (No.
2001CB309310).

\end{document}